\documentstyle[aps,twocolumn,epsbox]{revtex}
\begin{document}
\draft


\preprint{YITP-99-17, gr-qc/9904005}

\title{Hartle-Hawking state is a maximum of entanglement entropy} 
\author{Shinji Mukohyama}
\address{Yukawa Institute for Theoretical Physics, Kyoto University,
Kyoto 606-8502, Japan\\
Department of Physics and Astronomy, University of Victoria, 
Victoria BC, Canada V8W 3P6
}
\date{\today}
\maketitle

\begin{abstract}

It is shown that the Hartle-Hawking state of a scalar field is a
maximum of entanglement entropy in the space of pure quantum states 
satisfying the condition that backreaction is finite. 
In other words, the Hartle-Hawking state is a curved-space analogue of 
the EPR state, which is also a maximum of entanglement entropy.

\end{abstract}


Entanglement entropy~\cite{BKLS,Srednicki} is one of the strongest
candidates of the origin of black hole
entropy~\cite{Bekenstein,Hawking}. It is originated from a direct-sum 
structure of a Hilbert space of a quantum system: for an element
$|\psi\rangle$ of the Hilbert space ${\cal F}$ of the form 
%
\begin{equation}
 {\cal F} = {\cal F}_I \bar{\otimes} {\cal F}_{II},
	\label{eqn:F=F1*F2}
\end{equation}
the entanglement entropy $S_{ent}$ is defined by 
%
\begin{eqnarray}
 S_{ent} & = & -{\bf Tr}_I[\rho_I\ln\rho_I],\nonumber\\
 \rho_I  & = & {\bf Tr}_{II}|\psi\rangle\langle\psi|.
\end{eqnarray}
Here $\bar{\otimes}$ denotes a tensor product followed by a suitable
completion and ${\bf Tr}_{I,II}$ denotes a partial trace over 
${\cal F}_{I,II}$, respectively.

In Ref.~\cite{Mukohyama1998}, a new interpretation of the entanglement 
entropy was proposed based on its relation to the so-called
conditional entropy and a well-known meaning of the latter. 
It was conjectured that the entanglement entropy for a pure state is
an amount of information, which can be transmitted through 
${\cal F}_{II}$ and ${\cal F}_{I}$ from a system interacting with 
${\cal F}_{II}$ to another system interacting with ${\cal F}_{I}$ by
using quantum entanglement. 
Thus, it is important to seek quantum states having maximal
entanglement entropy and to investigate properties of the states. 
In fact, it was shown that a state having maximal entanglement entropy
plays an important role in quantum
teleportation~\cite{Mukohyama1998}.

In this letter, we show that the Hartle-Hawking
state~\cite{Hartle&Hawking} of a scalar field is a maximum of
entanglement entropy in the space of pure states satisfying a consistency
condition.

\begin{figure}
 \begin{center}
  \epsfile{file=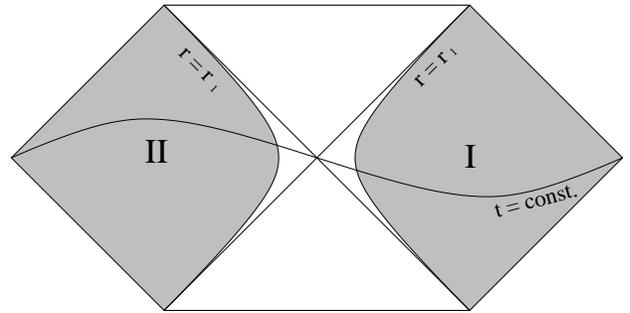,scale=0.4}
 \end{center}
\caption{
The Kruskal-like extension of the static, spherically symmetric
black-hole spacetime.
We consider only the regions satisfying $r>r_1$ (the shaded regions
$I$ and $II$ ).
}
\label{fig:Kruskal}
\end{figure}

For simplicity, we consider a minimally coupled, real scalar field
described by the action 
%
\begin{equation}
 S = -\frac{1}{2}\int d^4x\sqrt{-g}\left[
        g^{\mu\nu}\partial_{\mu}\phi\partial_{\nu}\phi
        + m^2\phi^2\right], 
\end{equation}
in the spherically symmetric, static black-hole spacetime
%
\begin{equation}
 ds^2 = -f(r)dt^2 + \frac{dr^2}{f(r)} + r^2d\Omega^2.
\end{equation}
We denote the area radius of the horizon by $r_0$ and the surface
gravity by $\kappa_0$ ($\ne 0$):
%
\begin{eqnarray}
 f(r_0) & = & 0,\nonumber\\
 \kappa_0 & = & \frac{1}{2}f'(r_0).
\end{eqnarray}
We quantize the system of the scalar field with respect to the Killing 
time $t$ in a Kruskal-like extension of the black hole spacetime. 
The corresponding ground state is called the Boulware state and its 
energy density is known to diverge near the horizon. 
Although we shall only consider states with bounded energy density, it
is convenient to express these states as excited states above the
Boulware ground state for technical reasons. 
Hence, we would like to introduce an ultraviolet cutoff $\alpha$ with 
dimension of length to control the divergence. Off course, we shall 
take the limit $\alpha\to 0$ in the end. 
The cutoff parameter $\alpha$ is implemented so that we only consider
two regions satisfying $r>r_1$ (shaded regions $I$ and $II$ in 
{\it Figure}~\ref{fig:Kruskal}), where $r_1$ ($>r_0$) is determined by 
%
\begin{equation}
 \alpha = \int_{r_0}^{r_1}\frac{dr}{\sqrt{f(r)}}. 
\end{equation}
(Strictly speaking, we also have to introduce outer boundaries, say at 
$r=L$ ($\gg r_0$), to control the infinite volume of the constant-$t$
surface. However, even if there are outer boundaries, the following
arguments still hold and we can take the limit $L\to\infty$ in the
end.) 
Evidently, the limit $\alpha\to 0$ corresponds to the limit 
$r_1\to r_0$. Thus, in this limit, the whole region in which 
$\partial /\partial t$ is timelike is considered. 

In this situation, there is a natural choice for division of the
system of the scalar field: let ${\cal H}_I$ be the space of mode
functions with supports in the region $I$ and ${\cal H}_{II}$ be the 
space of mode functions with supports in the region $II$. 
Thence, the space ${\cal F}$ of all states are of the form
(\ref{eqn:F=F1*F2}), where ${\cal F}_I$ and ${\cal F}_{II}$ are
defined as symmetric Fock spaces constructed from ${\cal H}_I$ and
${\cal H}_{II}$, respectively:
%
\begin{eqnarray}
 {\cal F}_I & \equiv &
    \mbox{\boldmath C} \oplus {\cal H}_I \oplus 
    \left({\cal H}_I\bar{\otimes} {\cal H}_I\right)_{sym}
    \oplus \cdots,      
    \nonumber   \\
 {\cal F}_{II} & \equiv &  
    \mbox{\boldmath C} \oplus {\cal H}_{II} \oplus 
    \left({\cal H}_{II} \bar{\otimes} {\cal H}_{II} \right)_{sym} 
    \oplus \cdots.
\end{eqnarray}
Here $(\cdots)_{sym}$ denotes the symmetrization.


Let us investigate what kind of condition should be imposed for our
arguments to be self-consistent. 
A clear condition is that the backreaction of the scalar field to the
background geometry should be finite.
Thus, the contribution $\Delta M$ of the subsystem ${\cal F}_I$ to
the mass of the background geometry should be bounded in the limit
$\alpha\to 0$. 
Although this condition is not a sufficient condition in order to make 
the backreaction finite, what is surprising is that we can obtain the
Hartle-Hawking state without using any other conditions stronger than
this. In this sense, our strategy of maximizing entanglement entropy
seems as strong as the maximum entropy principle in the statistical
thermodynamics.

As shown in Ref.~\cite{Mukohyama&Israel}, $\Delta M$ is given by
%
\begin{equation}
 \Delta M \equiv 
	-\int_{x\in I}T^t_t 4\pi r^2dr = H_I,
\end{equation}
where $H_I$ is the Hamiltonian of the subsystem ${\cal F}_I$ with
respect to the Killing time $t$.
Hence, the expectation value of $\Delta M$ with respect to a state
$|\psi\rangle$ of the scalar field is decomposed into the contribution 
of excitations and the contribution from the zero-point energy: 
%
\begin{equation}
 \langle\psi |\Delta M|\psi\rangle = E_{ent} + \Delta M_{B},
\end{equation}
where $E_{ent}$ is entanglement energy~\footnote{
This definition corresponds to $E_{ent}^{(I')}$ in Ref.~\cite{MSK1998} 
and $\langle :H_2:\rangle$ in Ref.~\cite{D-thesis}.
}
defined by
%
\begin{equation}
 E_{ent} \equiv \langle\psi |:H_I:|\psi\rangle, \label{eqn:Eent}
\end{equation}
and $\Delta M_{B}$ is the zero-point energy of the Boulware state. 
Here, the colons denote the usual normal ordering.

Since the Boulware energy $\Delta M_{B}$ diverges as 
$\Delta M_{B}\simeq -cAT_H\alpha^{-2}$ in the limit $\alpha\to 0$, we
should impose the condition  
%
\begin{equation}
 E_{ent} \simeq |\Delta M_B| \label{eqn:SBC}
\end{equation}
in the lowest order in $\alpha^{-1}$, where $A=4\pi r_0^2$ is the area of 
the horizon, $T_H=\kappa_0/2\pi$ is the Hawking temperature and $c$ is
a positive constant of order unity. We call this condition {\it the
small backreaction condition (SBC)}~\footnote{
Off course, finite energy can be added to the entanglement energy
without generating divergent backreaction. However, the finite energy
is higher order in the $\alpha^{-1}$ expansion. Thus, effects  of the 
finite energy are higher order in $\alpha^{-1}$. 
}.

Note that the right hand side of SBC (\ref{eqn:SBC}) is independent of
the state $|\psi\rangle$. 
Thus, SBC implies that the entanglement energy should be fixed when we
maximizes the entanglement entropy.  
In statistical thermodynamics, it is well known in what situation we
should fix average energy: we have to fix it when we know observed
value of energy. 
However, the corresponding situation seems not to have been
essentially known for (quantum or thermal) excitations of fields on a
black hole background. In fact, for the brick wall model, in which
thermal excitations are considered, Ref.~\cite{Mukohyama&Israel} is
the first which pointed out that the ground state for the brick wall
model is the Boulware state and that the Boulware (negative,
divergent) energy should be added to thermal energy. 
(See Refs.~\cite{tHooft,Liberati,Belgiorno&Martellini} for complete
confusion reigning on this subject.) 
For our system in this letter, it is the Boulware (negative,
divergent) energy that makes us fix the entanglement energy.


Now, we shall show that the Hartle-Hawking state is a maximum of the 
entanglement entropy in the space of pure quantum states satisfying
SBC. 
For this purpose, we prove a more general statement for a quantum
system with a state-space of the form (\ref{eqn:F=F1*F2}): 
{\it a state of the form 
%
\begin{equation}
 |\psi\rangle = {\cal N} \sum_n 
	e^{-E_n/2T}|n\rangle_{I}\otimes|n\rangle_{II}
	\label{eqn:HH-state}
\end{equation}
is a maximum of the entanglement entropy in the space of pure states
with fixed expectation value of the operator $E_{I}$ defined by 
%
\begin{equation}
 E_{I} = \left(\sum_n E_n|n\rangle_{I}\cdot{}_{I}\langle n|\right)
	\otimes 
	\left(\sum_m |m\rangle_{II}\cdot{}_{II}\langle m|\right),
\end{equation}
provided that the real constant $T$ is determined so that the
expectation value of $E_I$ is actually the fixed value. Here, 
${\cal F}_I$ and ${\cal F}_{II}$ are assumed to be isomorphic to each
other, $\{|n\rangle_I\}$ and $\{|n\rangle_{II}\}$ ($n=1,2,\cdots$) are
bases of the subspaces ${\cal F}_I$ and ${\cal F}_{II}$, respectively,
and $E_n$ are assumed to be real and positive.}
Note that this statement looks almost the same as the following
statement in statistical thermodynamics: a canonical state is a
maximum of statistical entropy in the space of states with fixed
energy, provided that the temperature of the canonical state is
determined so that the energy is actually the fixed value. Thus, it
might be expected that the above general statement might be directly
derived from the standard Jaynes method~\cite{Jaynes} as this
statement in statistical thermodynamics can be derived. However, the
Jaynes method cannot be applied directly to our system since Jaynes
method treats entropy of not a subsystem but a total system. Thus, in
the proof given below, we seek a correspondence between our
variational principle and the standard variational principle in
statistical thermodynamics. (See Eq.~(\ref{eqn:correspondence}).)

Note that the expectation value of $E_I$ is equal to the entanglement
energy (\ref{eqn:Eent}), providing that $|n\rangle_I$ and $E_n$ are an
eigenstate and an eigenvalue of the normal-ordered Hamiltonian $:H_I:$
of the subsystem ${\cal F}_I$. 
Hence, for the system of the scalar field, the above general statement
insists that the state (\ref{eqn:HH-state}) is a maximum of the
entanglement entropy in the space of pure states satisfying SBC, which 
corresponds to fixing the entanglement energy. 
Off course, in this case, the constant $T$ should be determined so
that SBC (\ref{eqn:SBC}) is satisfied.

Returning to the subject, let us prove the general statement. 
(The following proof is the almost same as that given in the Appendix
of Ref.~\cite{Mukohyama1998} for a slightly different statement. 
However, for completeness, we shall give the proof. )

First, we decompose an element $|\psi\rangle$ of ${\cal F}$ as 
%
\begin{equation}
 |\psi\rangle = \sum_{n,m}C_{nm}|n\rangle_I\otimes |m\rangle_{II}, 
\end{equation}
where the coefficients $C_{nm}$ ($n,m=1,2,\cdots$) are complex numbers 
satisfying $\sum_{n,m}|C_{nm}|^2=1$ and can be considered as matrix
elements of a matrix $C$. Since $C^{\dagger}C$ is a non-negative
Hermitian matrix, it can be diagonalized as
%
\begin{equation}
 C^{\dagger}C = V^{\dagger}PV,
\end{equation}
where $P$ is a diagonal matrix with diagonal elements $p_{n}$ 
($\ge 0$) and $V$ is a unitary matrix. 
For this decomposition and diagonalization, the entanglement
entropy and the expectation value of the operator $E_I$ are written as 
follows.
%
\begin{eqnarray}
 S_{ent} & = & -\sum_n p_n\ln p_n, \\
 E_{ent} & = & \sum_{n,m}E_n p_m|V_{nm}|^2, 
\end{eqnarray}
where $V_{nm}$ is matrix elements of $V$. The constraints
$\sum_{n,m}|C_{nm}|^2=1$ and $V^{\dagger}V={\bf 1}$ are equivalent to 
%
\begin{eqnarray}
 \sum_n p_n & = & 1,\nonumber\\
 \sum_l V^{*}_{ln}V_{lm} & = & \delta_{nm}.
\end{eqnarray}

Next, we shall show that these expressions are equivalent to those
appearing in statistical mechanics in ${\cal F}_I$. 
Let us consider a density operator $\bar{\rho}$ on ${\cal{F}}_I$: 
%
\begin{equation}
 \bar{\rho} = \sum_{n,m} \tilde{P}_{nm}
                |n\rangle_I\cdot {}_I\langle m|,
\end{equation}
where $(\tilde{P}_{nm})$ is a non-negative Hermitian matrix with unit
trace.  By diagonalizing the matrix $\tilde{P}$ as 
%
\begin{equation}
 \tilde{P} = \bar{V}^{\dagger}\bar{P}\bar{V},
\end{equation}
we obtain the following expressions for entropy $S$ and an expectation 
value $E$ of the operator
$\bar{E}_I\equiv\sum_n E_n|n\rangle_{I}\cdot{}_{I}\langle n|$. 
%
\begin{eqnarray}
 S & = & -\sum_n \bar{p}_n\ln \bar{p}_n,\nonumber\\
 E & = & \sum_{n,m} E_n\bar{p}_m |V_{nm}|^2,
\end{eqnarray}
where $\bar{p}_n$ is the diagonal elements of $\bar{P}$.
The constraints $\bf{Tr}\bar{\rho}=1$  and 
$\bar{V}^{\dagger}\bar{V}={\bf 1}$ are restated as 
%
\begin{eqnarray}
 \sum_n \bar{p}_n & = & 1,\nonumber\\
 \sum_l \bar{V}^{*}_{ln}\bar{V}_{lm} & = & \delta_{nm}.
\end{eqnarray}

From these and those expressions, the following correspondence is
easily seen:
%
\begin{eqnarray}
 S_{ent}	& \leftrightarrow & S,\nonumber\\
 E_{ent}	& \leftrightarrow & E,\nonumber\\
 C^{\dagger}C	& \leftrightarrow & \tilde{P}.
	\label{eqn:correspondence}
\end{eqnarray}
Hence, a maximum of $S$ in the space of statistical states with a
fixed value of $E$ gives a set of maxima of $S_{ent}$ in a space of
quantum states with a fixed value of $E_{ent}$. (All of them are
related by unitary transformations in the subspace ${\cal F}_{II}$.) 
Thus, since the thermal state $\tilde{P}_{nm}=e^{-E_n/T}\delta_{nm}$
is a maximum of $S$ in a space of statistical states with a fixed
value of $E$, $C_{nm}=e^{-E_n/2T}\delta_{nm}$ is a maximum of 
$S_{ent}$ in a space of pure quantum states with a fixed value of 
$E_{ent}$. Here the temperature $T$ should be determined so that $E$
(or $E_{ent}$) has the fixed value. This completes the proof of the
general statement.

Therefore, for the system of the scalar field, a state of the form
(\ref{eqn:HH-state}) is a maximum of the entanglement entropy in a
space of pure quantum states satisfying SBC, provided that the
constant $T$ is determined so that SBC is satisfied. The value of $T$
is easily determined as $T=T_H$ by using the well-known fact that the
negative divergence in the Boulware energy density can be canceled by
thermal excitations if and only if temperature with respect to the
time $t$ is equal to the Hawking temperature~\footnote{
See eg. Ref.~\cite{Mukohyama&Israel}.}.


Finally, by taking the limit $\alpha\to 0$, we obtain the statement
that the Hartle-Hawking state is a maximum of entanglement entropy in
a space of pure quantum states satisfying SBC~\footnote{
The Hartle-Hawking state is actually of the form (\ref{eqn:HH-state})
with $T=T_H$~\cite{Israel1976}, provided that $\alpha\to 0$.}.
In other words, the Hartle-Hawking state is a curved-space analogue of 
the EPR state, which is also a maximum of entanglement
entropy~\cite{Mukohyama1998}.

From this result we can say that the brick wall model of
'tHooft~\cite{tHooft} seeks the maximal value of entanglement
entropy~\cite{Mukohyama1999}. 
Thus, the maximal entanglement entropy is equal to the black hole
entropy if the cutoff length $\alpha$ is set to be of order of the
Planck length. Although in our arguments we have taken the limit
$\alpha\to 0$, it will be valuable to investigate possibilities that
quantum fluctuations of geometries may prevent $\alpha$ from being
zero and that the fluctuations of the horizon may be  
effectively represented as a Planck-order-value of $\alpha$. 
Note that the effect of non-zero value of $\alpha$ to our arguments
should be small enough if the mass of a background black hole is
sufficiently large in Planck unit.

Our arguments suggests strong connection among three kinds of
thermodynamics: black hole thermodynamics, statistical thermodynamics, 
and entanglement 
thermodynamics~\cite{Mukohyama1998,MSK1998,D-thesis,MSK1997}. 
Moreover, from the interpretation of entanglement entropy proposed in
Ref.~\cite{Mukohyama1998}, it is expected that the Hartle-Hawking
state may play an important role in transmitting information by using
quantum entanglement to restore temporarily missing information. It
will be interesting to investigate such a possibility in detail.

\vskip 1cm

\centerline{\bf Acknowledgments}
The author would like to thank Professor H. Kodama and Professor
W. Israel for their continuing encouragement.
This work was supported partially by the Grant-in-Aid for Scientific
Research Fund (No. 9809228).


\end{document}